\documentclass{aa}  

\usepackage{graphicx}
\usepackage{txfonts}
\usepackage{changes}
\definechangesauthor[color=violet]{SL}

\newcommand{\Uz}{m\,g\,z}

\begin{document} 
\title{Kinetic collisionless model of the solar transition region and corona with spatially intermittent heating}
\titlerunning{Kinetic model with spatially intermittent heating}
   \author{Luca Barbieri \thanks{\email{luca.barbieri@obspm.fr}}
          \inst{1}
          \and Pascal Démoulin
          \inst{1}
    }

   \institute{LIRA, Observatoire de Paris, 
     Université PSL, CNRS, Sorbonne Université, Université Paris-Cité, France
             }
   \date{Received xxx, yyyy; accepted xxx, yyyy}

\abstract{

\textit{Context.}
The solar corona exhibits a striking temperature inversion, with plasma temperatures exceeding $10^6$~K above a much cooler chromosphere. How the coronal plasma reaches such extreme temperatures remains a fundamental open question in solar and plasma physics, known as the coronal heating problem.

\textit{Aims.}
We investigate whether localized heating events, spatially distributed across the upper chromosphere and base of the transition region, combined with a collisionless corona, can self-consistently generate realistic temperature and density profiles without requiring direct energy deposition within the corona itself.

\textit{Models.}
We develop a three-dimensional kinetic model of a collisionless stellar atmosphere embedded in a uniform magnetic field, where heating occurs intermittently at the chromosphere–transition region interface. A surface coarse-graining procedure is introduced to capture the spatial intermittency of heating, leading to non-thermal boundary conditions for the Vlasov equation. We derive analytical expressions for the stationary distribution functions and compute the corresponding macroscopic profiles.

\textit{Results.}
We show that spatially intermittent heating, when coarse-grained over a surface containing many localized events, produces suprathermal particle distributions and a temperature inversion via velocity filtration. The resulting density and temperature profiles feature a transition region followed by a hot corona, provided that heating events are spatially sparse, consistently with solar observations. This result holds independently of the specific statistical distribution of temperature increments. Importantly, no local heating is applied within the corona.

\textit{Conclusions.}
The model demonstrates that spatial intermittency alone, i.e., a sparse distribution of heated regions at the chromospheric interface, is sufficient to explain the formation of the transition region and the high-temperature corona.
}

   \keywords{Sun: corona -- Sun: atmosphere -- plasmas -- methods: analytical}
   \maketitle

\section{Introduction}

The temperature profile of the Sun exhibits a notable reversal as a function of radial distance. Starting from the core, the temperature decreases outward through the radiative and convective zones, reaching a minimum at the photosphere. Beyond this point, in the low chromosphere, the temperature starts to gradually rise, reaching values around $10^4$ K. A dramatic increase then occurs across a remarkably thin layer known as the transition region, beyond which the temperature rapidly rises to exceed one million kelvin in the tenuous outer atmosphere, the solar corona. This abrupt change in the temperature gradient is referred to as the temperature inversion. Understanding the physical mechanisms responsible for heating the coronal plasma to such extreme temperatures remains one of the most fundamental open problems in solar and plasma physics, commonly known as the coronal heating problem.

Most classical models assume local thermodynamic equilibrium, implying that the corona must be directly heated from above \citep{Parker:1972wu,Ionson_1978,Heyvaerts_Priest_1983,Dmitruk:1997uf,2005ApJ...618.1020G,Klimchuk_2006,Rappazzo:2008vl,Reale2010,Dahlburg2012,2013ApJ...773L...2R,2015RSPTA.37340265W,Klimchuk2015,2020A&A...636A..40H,VanDoorsselaere2020,Viall2021}. However, observational evidence suggests that local thermodynamic equilibrium may not be valid in the transition region and corona \citep{Dudik:2017un}, allowing for alternative mechanisms.

An alternative class of solutions was proposed in the early 1990s \citep{Scudder1992a,Scudder1992b}, based on the idea that the coronal plasma may not be in thermal equilibrium. If one assumes the presence of suprathermal power-law tails in the velocity distribution functions (VDFs) of particles already in the upper chromosphere, then the hotter particles are more likely to escape the Sun’s gravitational potential. This results in a temperature that increases with height, in a mechanism known as velocity filtration or gravitational filtering. 
However, this interpretation faces two key limitations: first, it predicts a smooth variation of temperature and density, without a clearly defined transition region; second, the suprathermal tails required must exist at chromospheric heights where collisionality is strong and tends to enforce thermal equilibrium.

Still, while the chromospheric VDFs are likely to be close to thermal due to collisionality, the chromosphere itself is highly dynamic and structured, with fine-scale inhomogeneities revealed down to instrumental resolution \citep{Cauzzi:2009ta,Ermolli:2022um}. Observations and numerical simulations indicate that temperature can fluctuate significantly in space and time \citep{Peter:2014uz,Hansteen:2014us}.

Recently, \citet{barbieri2023temperature} introduced a kinetic N-particle model of coronal loops in the solar atmosphere. Using both numerical simulations and analytical analysis, \citet{Barbieri2024b,Barbieri2025b} show
that short-lived, intense, and temporally intermittent heating events in the upper chromosphere can drive the overlying collisionless plasma toward a steady-state configuration featuring a temperature inversion and a decreasing density profile similar to what is observed in the Sun’s atmosphere.

Because a million-degree corona is also present in low-mass main-sequence stars (i.e., stars with $M < 1.5 M_{\odot}$), and the velocity filtration mechanism is 
not Sun-specific, the same model was subsequently applied to other stars successfully predicting temperature inversion even in this case \citep{barbieri2024temperaturedensityprofilescorona}.

In those works, coronal loops were modeled as one-dimensional, unmagnetized, two-component collisionless plasmas subject to gravity and to an electrostatic field, in steady contact with a collisional chromosphere. The key result was that, in response to short and rapid heating pulses in the chromosphere, both shorter than the electron loop crossing time, suprathermal tails naturally develop in the overlying plasma, leading to velocity filtration and temperature inversion. Crucially, no heating was applied directly in the corona, and no non-thermal distributions were imposed at the base, in contrast to the assumptions of Scudder’s model.

In the present work, we introduce a three-dimensional kinetic model of a stellar atmosphere in which the coronal plasma is treated as collisionless, with the inclusion of magnetic field lines. Given that the chromosphere is a highly dynamic environment characterized by small-scale heating events distributed across the high chromosphere 
\citep{Dere:1989ux, Teriaca:2004wy, Peter:2014uz, Young2018-sp, Tiwari:2019us, Lee:ApJ2020, Berghmans:2021wl, Zhukov2021, Rauoafi:ApJ2023, Amari_2025,narang2025,Harra2025}, the thermal coupling with the corona 
is modelled via an interface surface, where localized heating events occur at discrete locations. Below, we demonstrate that small-scale, spatially distributed heating events can self-consistently reproduce the observed local density and temperature profiles, provided that they occupy only a small fraction of the total surface area.

Unlike \citet{barbieri2023temperature,Barbieri2024b,Barbieri2025b}, where temperature inversion emerged from temporal intermittency at a fixed location, the present model attributes the inversion to spatial intermittency, more precisely the temperature inversion is associated with the inhomogeneous coexistence of hot patches (with heating) and cold patches (without heating) along the chromosphere. While both approaches result in similar coarse-grained velocity distributions and inverted temperature profiles, the underlying mechanisms differ. The present model thus offers a complementary physical interpretation of transition region formation based on spatial intermittency.
Both temporal and spatial intermittences are simultaneously present in the solar atmosphere, so that they both contribute to the temperature inversion. Below, we isolate the spatial  intermittency to better characterise its properties.

The paper is structured as follows. In Sec. \ref{sec1}, we introduce the model. In Sec. \ref{sec2} we introduce the surface coarse-graining procedure and we derive an analytical expression for the particles distribution functions. In Sec.~\ref{sec3}, we establish the connection with the model presented in \citet{Barbieri2024b} and analyze the influence of the model parameters on the resulting temperature and density profiles, as well as on the corresponding particle velocity distribution functions. In Sec.  \ref{sec4}, we summarize the main results and outline possible future directions.

\begin{figure}
    \centering
    \includegraphics[width=0.99\columnwidth]{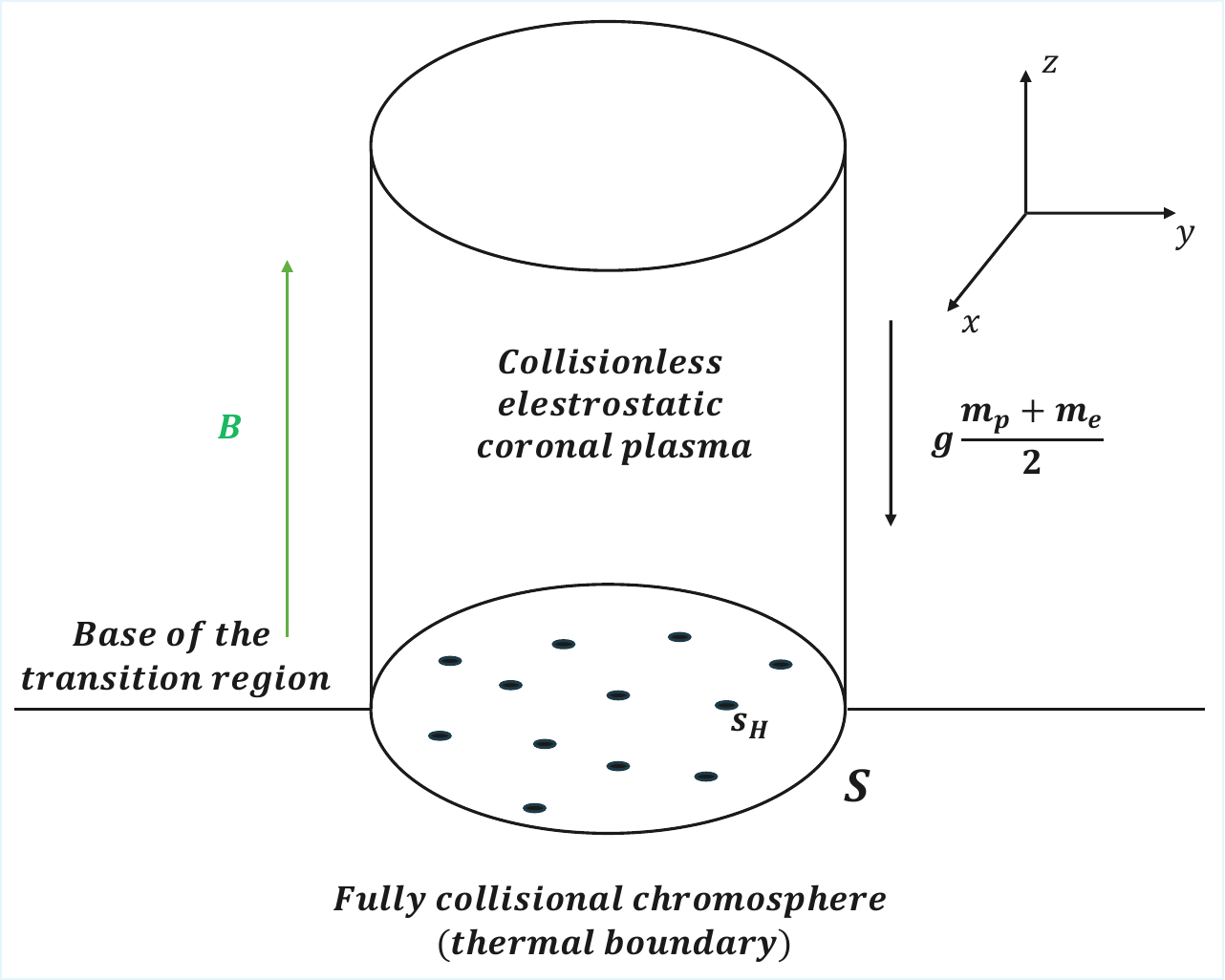}
    \caption{
    Schematic representation of the solar plasma model. The surface $S$, located at the base of the transition region, acts as the interface between the fully collisional chromosphere (serving as a thermal reservoir) and the collisionless coronal plasma. Localized heating events, each occupying an area $s_H$, are shown in black. The coronal plasma above is embedded in a uniform magnetic field $\textbf{B}$ (green), while particles are subject to a net external force $\textbf{g}(m_p + m_e)/2$ (black), which combines gravitational field and the Pannekoek–Rosseland electric field. The Cartesian reference frame $(x, y, z)$ used throughout the paper is indicated in the top right.}
    \label{fig:model}
\end{figure}

\section{The model}  
\label{sec1}

Motivated by the presence of numerous small-scale brightening events in the high chromosphere,
we consider a coarse-graining surface $S$ that satisfies the condition  
\begin{equation}\label{planeparallellimit}
    s_H \ll S \ll 4\pi R_{\odot}^2 \quad \mathrm{and} \quad n_H \gg 1 \quad,
\end{equation}
where $s_H$ denotes the characteristic size of a single heating event and $n_H$ is the number of such events contained within $S$. This condition ensures that $S$ is sufficiently large to statistically characterize the heating process, while still small enough compared to the full solar surface to retain a spatial variation of the parameters at larger scales (like coronal loops).
A schematic representation of the model setup is shown in Fig.~\ref{fig:model}.

Based on Eq.~\eqref{planeparallellimit}, we adopt a plane-parallel approximation and model a vertical slab of plasma 
located above the surface $S$. The plasma is treated as a collisionless, two-species system consisting of electrons and protons and we impose local charge neutrality. In addition, particles are subject to external forces: gravitational acceleration and the Pannekoek–Rosseland (PR) electric field, both resulting from the Sun's mass and charge 
neutralisation processes \citep{Pannekoek_1922, Rosseland_1924, Neslusan2001-rp, belmont2013collisionless,Barbieri_2025c}. The total external force acting on a particle of species \( \alpha \in \{e, p\} \) is  
\begin{equation}
    \textbf{F}_{\mathrm{ext},\alpha} = m_{\alpha} \textbf{g} + e_{\alpha} \textbf{E}_{\mathrm{PR}}  = \frac{(m_p +m_e)\,  \textbf{g} }{2} = m \,  \textbf{g}
    \quad,
\end{equation}
where $m$ is the mean particle mass and the PR electric field is given by  
\begin{equation}
     \textbf{E}_{\mathrm{PR}} = -\frac{m_p - m_e}{2 e} \textbf{g}\quad.
\end{equation}

Here, $\textbf{g} = g\hat{z}$ is the gravitational acceleration at chromospheric heights, with $g = GM_{\odot}/R_{\odot}^2$, and $\hat{z}$ is the unit vector pointing outward (away from the Sun). Since for simplicity we consider $g$ to be constant, this approach is valid for coronal plasma extending upward by a small fraction of the solar radius. We denote $m_e$ and $m_p$ as the electron and proton masses, respectively, and adopt the standard charge convention: $e_{\alpha} = +e$ for protons and $e_{\alpha} = -e$ for electrons. The system is immersed in a uniform magnetic field $\textbf{B}$ aligned with the $\hat{z}$ direction.

Under these assumptions, the distribution function $f_{\alpha}$ evolves according to the Vlasov equation:
\begin{equation}\label{Vlasovdynamics}
    \frac{\partial f_{\alpha}}{\partial t} + \textbf{v} \cdot \nabla f_{\alpha} + \frac{\textbf{F}_{\alpha}}{m_{\alpha}} \cdot \nabla_{\textbf{v}} f_{\alpha} = 0 \quad,
\end{equation}
where the total force acting on species $\alpha$ is  
\begin{equation}\label{totalforce}
    \textbf{F}_{\alpha} = 
       m\, \textbf{g} 
      + \frac{e_{\alpha}}{c} \textbf{v} \times \textbf{B} \quad,
\end{equation}

The plasma is in thermal contact with a boundary at the base $z = 0$, which represents the fully collisional chromosphere. This boundary is modeled as a Maxwellian reservoir described by an isotropic particle distribution
\begin{equation}\label{maxwellianboundary}
    f_{T_0,\alpha} (v)= n_0\left(\frac{m_{\alpha}}{2\pi k_B T_0}\right)^{3/2} e^{-\frac{m_{\alpha}v^2}{2k_B T_0}} \quad,
\end{equation}
where $n_0$ and $T_0$ are the local number density and temperature at a given point on the surface $S$ (i.e., in the high chromosphere), and $v$ is the particle velocity. 

Observational evidence indicates that the upper chromosphere and the base of the transition region are highly dynamic, exhibiting frequent, localized, and short-lived heating events \citep{Dere:1989ux, Teriaca:2004wy, Peter:2014uz, Young2018-sp, Tiwari:2019us, Lee:ApJ2020, Berghmans:2021wl, Zhukov2021, Rauoafi:ApJ2023, Nelson2024, Amari_2025,narang2025}. For an extended review see \cite{Harra2025} and references therein. As a result, the effective boundary temperature at $z = 0$ is expected to fluctuate spatially across $S$. We model this stochastic heating as follows:
\begin{itemize}
    \item Within a fraction of the surface $s_H$, localized heating events raise the temperature to $T = T_0 + \Delta T$, with $T$ drawn from a probability distribution $\gamma(T)$.
    \item Outside these regions, the temperature remains at the chromospheric background value $T_0$.
\end{itemize}

\section{Surface coarse-graining}\label{sec2}

We now introduce a coarse-grained description of the Vlasov dynamics based on surface-averaging, inspired by a recently proposed temporal coarse-graining method \citep{Barbieri2024b,Barbieri2025b}.

\subsection{Coarse-grained dynamics of the coronal plasma}

The surface coarse-grained phase-space distribution functions are defined as spatial averages of $f_{\alpha}$ over the interface surface $S$, namely
\begin{equation}
    \tilde{f}_{\alpha} = \langle f_{\alpha} \rangle_{S} 
    = \frac{1}{S} \int_{S} f_{\alpha} \, dx\, dy \quad.
\end{equation}
More generally, the coarse-grained version of any function $f$ of the phase-space coordinates is defined as
\begin{equation}
    \tilde{f} = \langle f \rangle_{S} 
    = \frac{1}{S} \int_{S} f \, dx\, dy \quad.
\end{equation} 

Applying the surface average to the Vlasov equation \eqref{Vlasovdynamics}, and noting that $\mathbf{F}_{\alpha}[f_{\alpha}]$ in Eq.~\eqref{totalforce} depends only on $\mathbf{v}$ and is therefore unaffected by the averaging, we obtain
\begin{equation}\label{coarsegraineddynamicswithboundary}
   \frac{\partial \tilde{f}_{\alpha}}{\partial t} 
    + v_z \frac{\partial \tilde{f}_{\alpha}}{\partial z} 
    + \frac{\mathbf{F}_{\alpha}}{m_{\alpha}} 
    \cdot \nabla_{\mathbf{v}} \tilde{f}_{\alpha} 
    = -\frac{1}{S}\int_{S} \mathbf{v}_{\perp} \cdot \nabla_{(x,y)} f_{\alpha}\, dx\,dy \quad,
\end{equation}
where $v_z = \mathbf{v}\cdot \hat{z}$ and $\mathbf{v}_{\perp}$ is the component of $\mathbf{v}$ orthogonal to $\hat{z}$.  

Since $\mathbf{v}_{\perp}$ does not depend on $x$ and $y$, the Green–Gauss theorem gives
\begin{equation}\label{boundaryterm}
    \int_{S} \mathbf{v}_{\perp} \cdot \nabla_{(x,y)} f_{\alpha}\, dx\,dy 
    = \mathbf{v}_{\perp}\cdot \int_{\partial S} \hat{n}(l)\,f_{\alpha}\,   dl \quad,
\end{equation}
where $\hat{n}(l)$ is the unit vector orthogonal to $\partial S$ in the $(x,y)$ plane, pointing outward from the surface $S$, and $dl$ is the infinitesimal line element along the curvilinear coordinate $l$ of the contour $\partial S$. For any surface $S$ whose boundary $\partial S$ does not intersect a heating event, Eq.~\eqref{boundaryterm} vanishes because $f_{\alpha}$ is constant along $\partial S$, and $\int_{\partial S} \hat{n}(l)\, dl = 0$.

Under these general conditions, Eq.~\eqref{coarsegraineddynamicswithboundary} reduces to
\begin{equation}\label{coarsegraineddynamicsinitial}
   \frac{\partial \tilde{f}_{\alpha}}{\partial t} 
    + v_z \frac{\partial \tilde{f}_{\alpha}}{\partial z} 
    + \frac{\mathbf{F}_{\alpha}}{m_{\alpha}} 
    \cdot \nabla_{\mathbf{v}} \tilde{f}_{\alpha} = 0 \quad.
\end{equation}

Since $\tilde{f}_{\alpha}$ depends only on $z$, this equation can equivalently be written as
\begin{equation}\label{coarsegraineddynamics}
   \frac{\partial \tilde{f}_{\alpha}}{\partial t} 
    + \mathbf{v} \cdot \nabla \tilde{f}_{\alpha} 
    + \frac{\mathbf{F}_{\alpha}}{m_{\alpha}} 
    \cdot \nabla_{\mathbf{v}} \tilde{f}_{\alpha} = 0 \quad.
\end{equation}
so that as the classical collision-less Vlasov equation.

In summary, the coarse-grained distribution functions $\tilde{f}_{\alpha}$ still satisfy Vlasov-type equations.

\subsection{Coarse-grained boundary conditions and stationary state}

To extract the coarse-grained distribution function at the boundary, $z=0$, we average the particle distribution $f_{T,\alpha}$ across the surface $S$:
\begin{equation}
    \tilde{f}_{\alpha}(0,v) = 
      A\, \langle f_{T,\alpha}(0,v) \rangle_{\sum_{i=1}^{n_H}  s_{H,i}} 
    + (1 - A)\, f_{T_0,\alpha}(0,v) \quad,
\end{equation}
where $A$ denotes the fraction of the surface subject to heating, defined as
\begin{equation}
    A = \frac{\sum_{i=1}^{n_H}  s_{H,i}}{S} \quad.
\end{equation}
In the limit $n_H \gg 1$, we invoke ergodicity and replace the spatial average with an ensemble average over the temperature distribution $\gamma(T)$:
\begin{equation} \label{eq:f_brightenings}
    \langle f_{T,\alpha}(0,v) \rangle_{\rm \sum_{i=1}^{n_H}  s_{H,i}} 
    = \int_{T_0}^{\infty} \gamma(T)\, f_{T,\alpha}(v) \, dT \quad,
\end{equation}
where the Gaussian function $f_{T,\alpha}(v)$ is defined as Eq.~\eqref{maxwellianboundary} with $T$ replacing $T_0$ and $\gamma(T)$ is a probability distribution so that $\int_{T_0}^{\infty} \gamma(T)\, dT =1$. This distribution function $\langle f_{T,\alpha}(0,v) \rangle_{n_H \cdot s_H}$ of the brightening events is defined by the physical processes involved such as the distribution of the electric field in the reconnection regions or the strength and inclination on the local magnetic field of the shocks, then averaged over a large number of brightening events. Here, we suppose that it can be decomposed in Gaussians centred on $v=0$, so that  $\langle f_{T,\alpha}(0,v) \rangle_{n_H \cdot s_H}$ is symmetric and a decreasing function of $v$ away from $v=0$ (with $\gamma(T)\geq 0$). Equation~\eqref{eq:f_brightenings} represents a large variety of velocity distribution with a concentrated core (with a thin limit fixed by $T_0$) and extended wings, monotonously decreasing with $v$ but otherwise with quite general shapes (defined by $\gamma(T)$).  Finally, within the above limits, the distribution function $\gamma(T)$ is determined by the acceleration processes involved in brightening events and we explore below very different shapes. 

The above procedure yields a compact expression for the coarse-grained distribution function at the base ($z=0$):
\begin{equation}\label{coarsegrainedflux}
    \tilde{f}_{\alpha}(0,v) = \mathcal{N}_{\alpha} \left[ A 
      \int_{T_0}^{\infty} \frac{\gamma(T)}{T^{3/2}} e^{-\frac{m_{\alpha}v^2}{2k_B T}} \, dT 
      + \frac{1 - A}{T_0^{3/2}} e^{-\frac{m_{\alpha}v^2}{2k_B T_0}} \right] 
\end{equation}
where the normalization constant $\mathcal{N}_{\alpha}$ is given by
\begin{equation}
    \mathcal{N}_{\alpha} = n_0\left(\frac{m_{\alpha}}{2\pi k_B}\right)^{\frac{3}{2}} \quad,
\end{equation}

Since the coarse-grained dynamics described by Eq.~\eqref{coarsegraineddynamics} remains Vlasov-like, the stationary solution can be obtained by applying Liouville’s theorem together with the boundary conditions. This yields
\begin{equation}\label{eq:falphastationary}
    \tilde{f}_{\alpha}(z,v) = \mathcal{N}_{\alpha} \left[ 
    A \int_{T_0}^{\infty} \frac{\gamma(T)}{T^{3/2}} 
    e^{-\frac{\mathcal{H}_{\alpha}}{k_B T}} dT 
    + \frac{1 - A}{T_0^{3/2}} 
    e^{-\frac{\mathcal{H}_{\alpha}}{k_B T_0}}
    \right] \quad,
\end{equation}
where the single-particle energy $\mathcal{H}_{\alpha}$ is given by
\begin{equation}
    \mathcal{H}_{\alpha} = \frac{1}{2} m_{\alpha} v^2 + m\,g\,z \quad,
\end{equation}

Equation~\eqref{eq:falphastationary} 
demonstrates that the steady-state distribution consists of a core Maxwellian component at temperature $T_0$, which decreases with height due to gravity, together with a suprathermal tail generated by spatially localized heating. As the height increases, the tail becomes increasingly dominant, giving rise to a temperature inversion through velocity filtration.

In the limiting case $A \rightarrow 0$, i.e. in the absence of heating events, the system relaxes to thermal equilibrium at the chromospheric temperature so with $\tilde{f}_{\alpha}$ given by Eq.~\eqref{maxwellianboundary}.

\begin{figure}
    \centering
   \includegraphics[width=0.99\columnwidth]{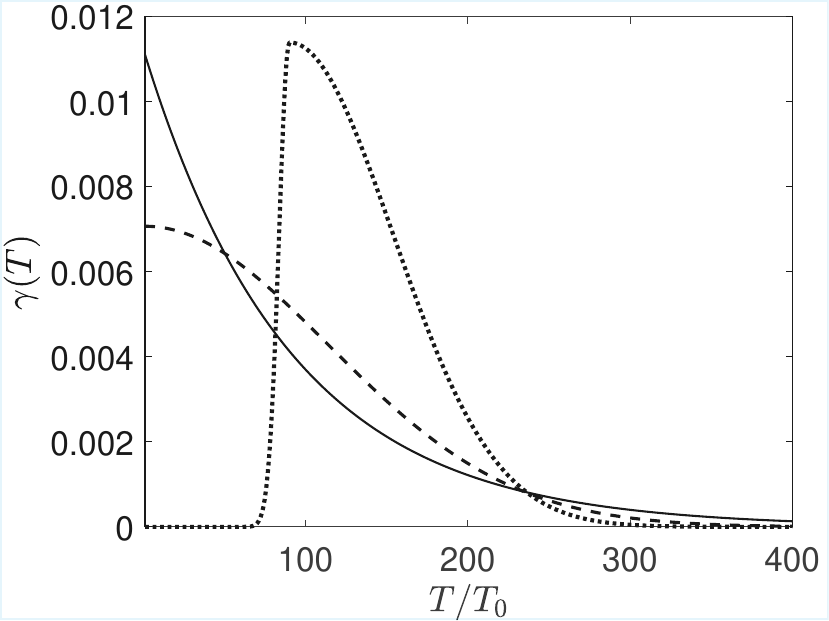}
    \caption{The three probability distribution functions $\gamma(T)$ used in next figures.  $\gamma_1(T)$ is defined by Eq.~\eqref{exponentialincrements} (continuous line), $\gamma_2(T)$ by Eq.~\eqref{halfgaussianincrements} (dashed line), and $\gamma_3(T)$ by Eq.~\eqref{twosidedgaussian} (dotted line). The parameters are
    $T_0 = 10^4$~K, and $\Delta T = 90\, T_0$. For the distribution $\gamma_3(T)$, we choose $T_h = \Delta T = 90\, T_0$, with $T_R = T_h$ and $T_L = 0.1\, T_h$ to satisfy the observational constraints. 
}
    \label{fig:gammas}
\end{figure}

\begin{figure*}
    \centering
    \includegraphics[width=0.99\textwidth]{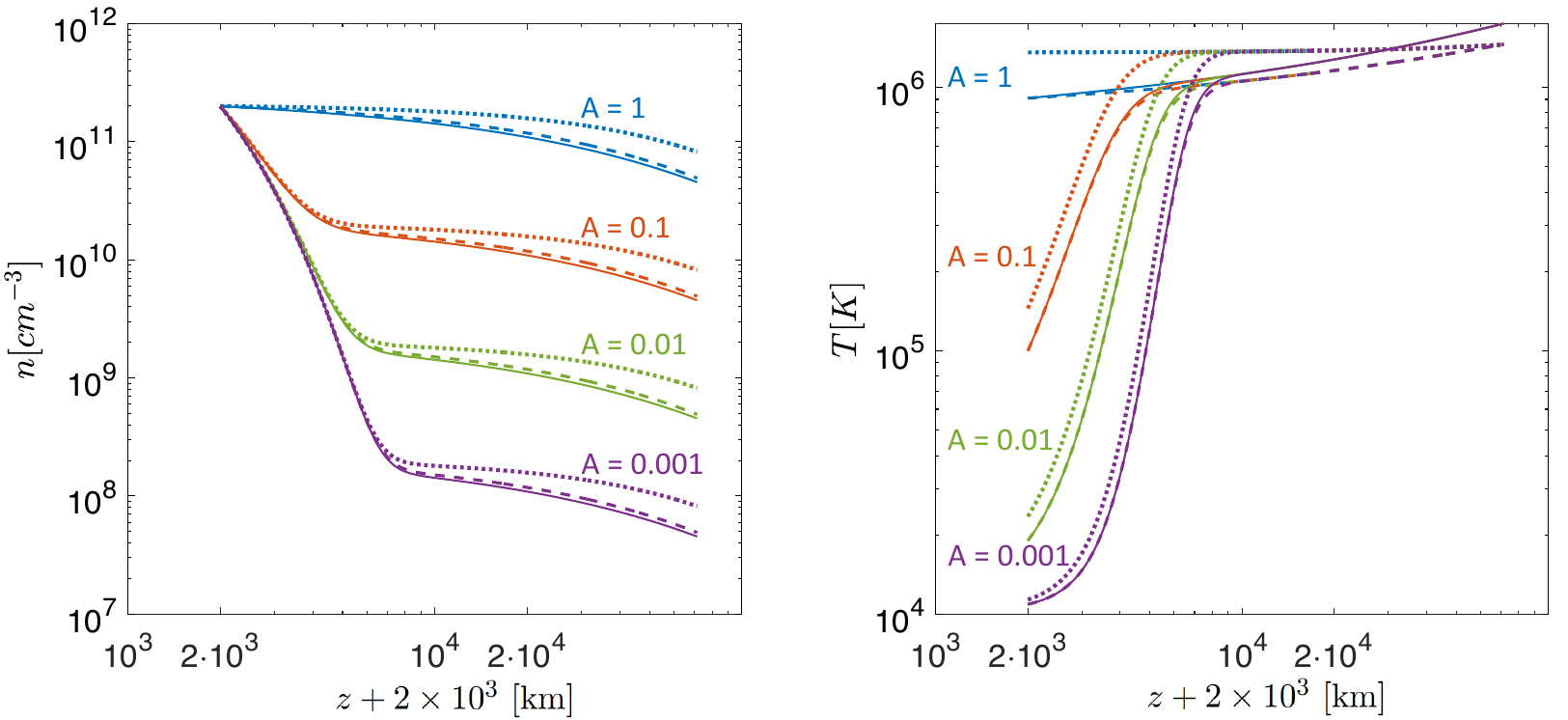}
    \caption{
    Right panel: number density profiles $\mathrm{cm}^{-3}$ as function of the height expressed in km computed using the distribution of heating events $\gamma_1(T)$ given by Eq. \eqref{exponentialincrements} (solid), 
    $\gamma_2(T)$ given by Eq. \eqref{halfgaussianincrements} (dashed) 
    and $\gamma_3(T)$ given by Eq. \eqref{twosidedgaussian} (dotted).  
    Blue lines correspond to $A = 1$, red lines to $A = 0.1$, green lines to $A = 0.01$ and purple lines to $A=0.001$. 
    Right panel: temperature profiles $T[K]$ as functions of the height expressed in km. The profiles are computed for the same distribution of temperature increments and values of $A$ of the left column. Moreover the same color coding and line style has been used.
    }
    \label{fig:temperatureinversion}
\end{figure*}

\begin{figure*}
    \centering
    \includegraphics[width=0.99\textwidth]{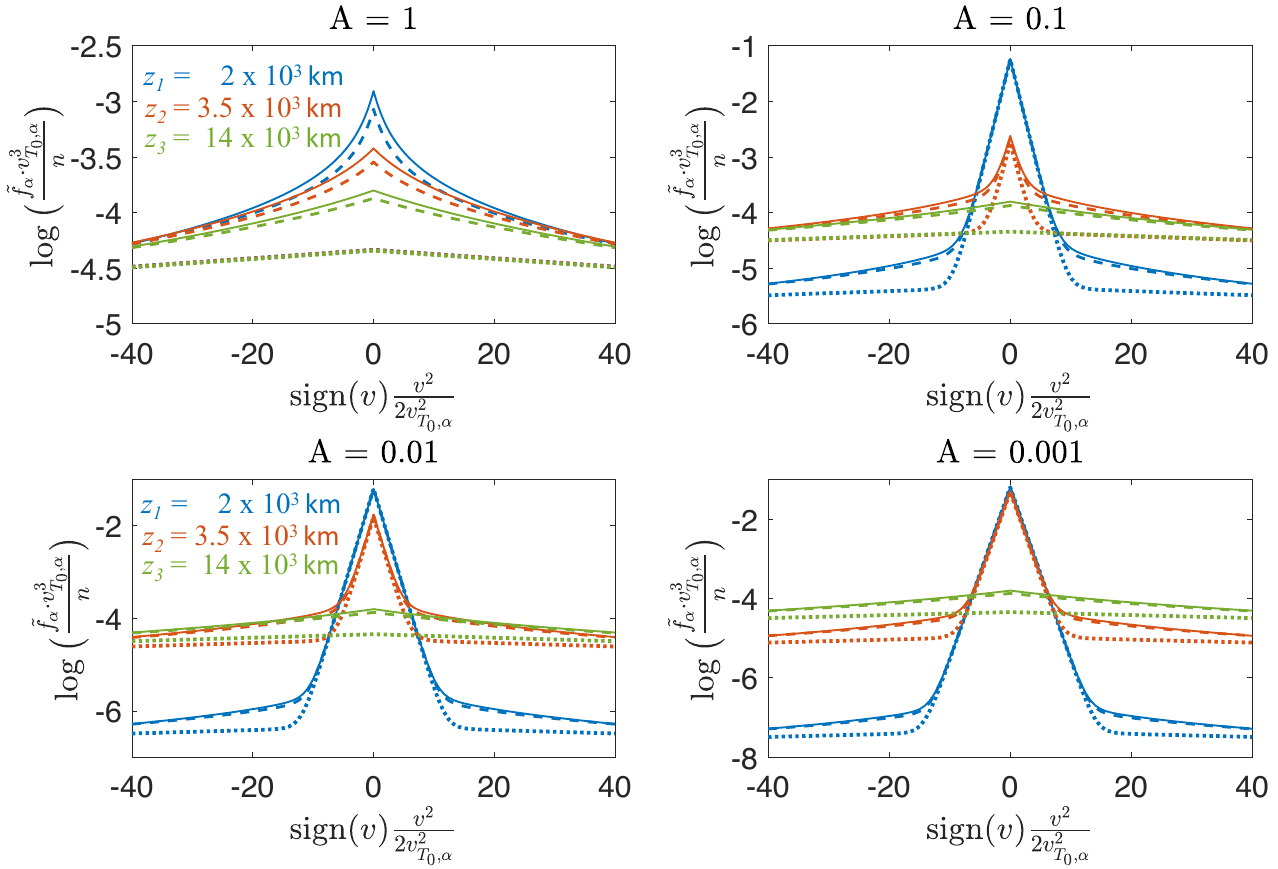}
    \caption{ 
    Decimal logarithm of the velocity distribution functions (VDFs), for species $\alpha$,  plotted as a function of the signed kinetic energy normalized by $v_{T_0,\alpha}^2$. The VDFs are scaled by the corresponding number densities 
    and normalized by $v_{T_0,\alpha}^3$, where $v_{T_0,\alpha}$ is defined by Eq.~\eqref{v_T_0}. Each panel compute the VDFs for different values of $A$ as shown in the subplot title. In each panel the VDFs are computed using the three distribution of temperature increments $\gamma_1(T)$ defined in Eq.\eqref{exponentialincrements} (solid),  
    $\gamma_2(T)$ defined in Eq.\eqref{halfgaussianincrements} (dashed) 
    and $\gamma_3(T)$ defined in Eq.\eqref{twosidedgaussian} (dotted).  
    Finally, in each panel, the VDFs are shown at three different heights (see their positions in Fig. \ref{fig:temperatureinversion}): $z_1 = 2 \times 10^3$ km (base location, blue), $z_2 = 3.5 \times 10^3$ km (transition region, red), and $z_3 = 14 \times 10^3$ km (corona, green).  
    }
    \label{fig:VDFs}
\end{figure*}

\section{Temperature, density, and particle distribution function results}
\label{sec3}

\subsection{Generic temperature and density profiles}

The full phase-space distribution functions derived here are mathematically analogous to Eq. (3.14) in \citet{Barbieri2024b}.
Moreover, although our model is based on a plane-parallel atmospheric geometry, Eq.~\eqref{eq:falphastationary} can be naturally extended to curved geometries by parametrizing the vertical coordinate $z$ along a curvilinear arc length, following the approach in \citet{Barbieri2024b}.

A further distinction lies in the interpretation of the parameter $A$: in the present model, $A$ represents the fraction of surface area undergoing heating events, whereas in \citet{Barbieri2024b}, $A$ denotes the fraction of time during which heating is active at the chromospheric boundary.

Using standard kinetic theory, one can compute the temperature and density profiles. Specifically, the number density is given by
\begin{equation}\label{density}
n(z) = n_{0} \left[\ A\, 
     \int_{T_0}^{\infty} \gamma(T)\, e^{-\frac{\Uz}{k_B T}} dT 
   + (1 - A)\, e^{-\frac{\Uz}{k_B T_0}} 
              \right]\quad,
\end{equation}
while the kinetic temperature profile is
\begin{equation}\label{temperature}
T(z) = \frac{ 
      A\, \int_{T_0}^{+\infty} T\,\gamma(T)\, e^{-\frac{\Uz}{k_B T}}dT 
    + (1-A)\, T_0\, e^{-\frac{\Uz}{k_B T_0}}
             }{
     A\, \int_{T_0}^{+\infty}\gamma(T)\, e^{-\frac{\Uz}{k_B T}}dT 
     + (1-A)\, e^{-\frac{\Uz}{k_B T_0}}
            } \quad.
\end{equation}

The constraints on the stochastic heating parameters identified in \citet{Barbieri2024b} remain valid in this framework:
\begin{itemize}
    \item $A \ll 1$ (i.e., $s_H \ll S$), ensuring $T(z = 0) \approx T_0$ at the base of the transition region, which implies that heating events are spatially sparse;
    \item $\Delta T = \int_{T_0}^{+\infty} (T - T_0)\gamma(T)\, dT \approx 10^2 T_0$, which is required to sustain a coronal temperature of approximately $10^6$~K given $T_0 \approx 10^4$~K.
\end{itemize}

When these conditions are met, the model reproduces realistic temperature and density profiles featuring a transition region followed by a hot, extended corona.
More precisely:
\begin{equation}\label{twosidedgaussian}
\begin{cases}
T(z) \approx T_0 
             & \text{for } z << \frac{k_B T_0}{m\,g} \\
T(z) \approx A\, \int_{T_0}^{+\infty} T\,\gamma(T)\, 
            e^{\frac{\Uz}{k_B} (-\frac{1}{T} +\frac{1}{T_0})}dT 
             & \frac{k_B T_0}{m\,g} << z << \frac{k_B \Delta T}{m\,g} \\
T(z) \approx \frac{ 
      \int_{T_0}^{+\infty} T\,\gamma(T)\, e^{-\frac{\Uz}{k_B T}}dT 
             }{
      \int_{T_0}^{+\infty}\gamma(T)\, e^{-\frac{\Uz}{k_B T}}dT 
            }
            & \frac{k_B \Delta T}{m\,g} << z 
\end{cases} 
\end{equation}
Below, we explicitly show that this outcome is weakly dependent of the specific form of the distribution $\gamma(T)$.

\subsection{Specific cases}
A fully constrained expression for the observational temperature increment distribution $\gamma(T)$ at the base of the transition region is currently lacking. Therefore, we consider three representative cases to explore a range of plausible scenarios. 

\paragraph{1. Exponential distribution:}
\begin{equation}\label{exponentialincrements}
\gamma_1(T) = \frac{1}{\Delta T} e^{-\frac{T - T_0}{\Delta T}} \quad \text{for } T > T_0 \quad.
\end{equation}

\paragraph{2. One-sided Gaussian distribution:}
\begin{equation}\label{halfgaussianincrements}
\gamma_2(T) = \frac{2}{\pi \Delta T} e^{-\frac{(T - T_0)^2}{\pi (\Delta T)^2}} \quad \text{for } T > T_0 \quad.
\end{equation}

Both $\gamma_1$ and $\gamma_2$ are monotonically decreasing in $T$, meaning that small temperature increments are more probable than large ones. These distributions represent the least favorable conditions for heating the corona, yet they still satisfy $\Delta T \gg T_0$.

\paragraph{3. Two-sided Gaussian distribution:}
\begin{equation}\label{twosidedgaussian}
\gamma_3(T) = 
\begin{cases}
C \cdot e^{-\frac{(T - T_h)^2}{T_R^2}} & \text{for } T \geq T_h \\
C \cdot e^{-\frac{(T - T_h)^2}{T_L^2}} & \text{for } T < T_h
\end{cases} \quad,
\end{equation}
where $C$ is set by $\int_{T_0}^{\infty} \gamma_3(T)\, dT =1$, which implies
\begin{equation}
   C = \frac{2}{\sqrt{\pi}} \frac{1}{T_R + T_L \mathrm{erf}\left(\frac{Th-T_0}{T_L}\right)} \quad.
\end{equation}
Here, $T_h$ denotes the peak temperature of the distribution, $T_R$ controls the spread above $T_h$, and $T_L$ governs the spread below it. This distribution is motivated by observational evidence indicating that heating events below $10^6$ K are rare \citep{Parker1988,Hudson1991,Parnell_2000,Bingert2013,Berghmans:2021wl,Purkhart2022,narang2025}, suggesting $T_L \ll T_h$. At the same time, heating events exceeding $T_h$ are observed, though they occur less frequently, justifying $T_R \sim T_h$ and the requirement for $\gamma_3(T)$ to decrease monotonically for $T > T_h$.

The selected $\gamma (T)$ distributions are shown in Figure~\ref{fig:gammas} with parameters selected to represent typical observations.

\subsection{Temperature and density profiles}

Figure~\ref{fig:temperatureinversion} shows the resulting temperature and density profiles for electrons and protons, which are identical in the stationary state due to the shared boundary condition and electric neutralisation (same density). 
The blue, red, green, and purple curves correspond to $A = 1$, $A = 0.1$, $A = 0.01$, and $A = 0.001$, respectively.

The left panel displays the density profiles, while the right panel shows the corresponding temperature profiles. For each value of $A$, results are plotted for all three $\gamma(T)$ distributions: $\gamma_1(T)$ (solid line), $\gamma_2(T)$ (dashed), and $\gamma_3(T)$ (dotted). The profiles show excellent agreement across the different forms of $\gamma(T)$.

Reducing $A$ results in the emergence of a transition region followed by a hot corona. For example, at $A = 0.001$, the temperature rises from $1.2 \times 10^4$~K to $5 \times 10^5$~K between $z = 2000$ and $5000$~km, while the density drops by two orders of magnitude.

On the right panel of Figure~\ref{fig:temperatureinversion}, the density profiles exhibit an opposite trend than temperature profiles: the density drops rapidly across the transition region and then more gradually in the corona. This behavior arises from the different gravitational scale heights associated with the thermal and suprathermal populations:
\begin{equation}
z_{T_0} = \frac{k_B T_0}{m\, g}\qquad \text{and} \qquad
z_{\Delta T} =
\frac{k_B (T_0+\Delta T)}{m\, g}  \quad,
\end{equation}
with $z_{\Delta T} \gg z_{T_0}$ due to $\Delta T \sim 100\, T_0$. As $A$ decreases, the cold population with a short scale height begins to dominate at low heights ($z<<z_{T_0}$). As $z$ is comparable to $z_{T_0}$, a rapid drop in density occurs above the base. In contrast, the suprathermal component, characterized by a larger scale height, becomes dominant in the corona, resulting in a more gradual density decrease at higher heights.

The $\gamma(T)$ functions selected have very different profiles (Fig. \ref{fig:gammas}). These differences have only a small effect on $n(z)$ and $T(z)$ (Fig. \ref{fig:temperatureinversion}). Even with $\gamma_3(T)$, which mostly includes coronal temperatures, the implied increase of $n(z)$ and $T(z)$ is limited. More noticable, the transition from chromosphere top to corona has nearly the same shape as for $\gamma_1(T)$ and $\gamma_2(T)$.

The transition region starts when the hot component starts to dominates the cold one, so the temperature rise. Then, we define the base of the transition region, at $z=z_B$, when the two terms in the numerator of Eq.~\eqref{temperature} are equal. In the same way, we define the top of the transition region, at $z=z_T$, when the two terms in the denominator of Eq.~\eqref{temperature} are equal, so the stabilisation of the temperature (only slightly increasing in the coronal part).  Since the transition region is thin compared to the gravitationnal scale height of the corona, for both height estimation we can use the simplification $e^{-\Uz / k_B T} \approx 1$ for the hot component.  These approximations provide
\begin{align}
  z_B &\approx z_{T_0} \,\,
         \log \left( \frac{1-A}{A} 
                      \frac{T_0}{\int_{T_0}^{+\infty} T\,\gamma(T)\, dT} 
            \right) \\
  z_T &\approx z_{T_0} \,\,
         \log \left( \frac{1-A}{A} 
            \right)  \quad.
\end{align}
This shows that $z_B$ depends of $\gamma(T)$ only with the average temperature while $z_T$ is not dependent on $\gamma(T)$. The effect of $A$ is only to shift the transition region globally in height.  The transition region thickness is simply
\begin{equation} \label{TR_thickness}
z_T-z_B \approx z_{T_0} \,\,
         \log \left( \frac{\int_{T_0}^{+\infty} T\,\gamma(T)\, dT}{T_0} 
            \right) \quad,
\end{equation}
so it is mainly defined by the gravitational scale height at $T_0$ and it weakly depends on $\gamma(T)$ (only a logarithm dependence on the mean temperature).

The width of the transition region in our model ($\sim 3000$ km) is broader than classical hydrodynamic or MHD estimates \citep[$< 200$~km, ][]{Klimchuk_2006}, although both the shape and amplitude of the profiles remain consistent with observational data \citep{GolubPasachoff:book,observedtemperature}.

\subsection{Velocity distribution functions}

Figure~\ref{fig:VDFs} shows the velocity distribution functions (VDFs), divided by $n(z)$, plotted as a function of the signed and normalized kinetic energy, $\mathrm{sign}(v)\, v^2/(2 v_{T_0,\alpha}^2)$, in semi-log scale, with the thermal speed
  \begin{equation}\label{v_T_0}
  v_{T_0,\alpha} = \sqrt{k_B T_0 / m_{\alpha}} \quad.
  \end{equation} 
In this representation, thermal (Gaussian) distributions appear as symmetric triangles centered at zero.

The qualitative trends are consistent across the three choices of $\gamma(T)$, as follows. At low heights, ($z<<z_{T_0}$), and for all values of $A$, the VDF exhibits a Maxwellian core centered at zero velocity, whose amplitude decreases with height on a scale height $z_{T_0}$. This component corresponds to the thermal background in Eq.~\eqref{eq:falphastationary}. 

At greater heights, the suprathermal tails, arising from the non-thermal component of the distribution, become increasingly dominant due to velocity filtration \citep{Scudder1992a,Scudder1992b}. This gravitational filtering progressively removes low-energy particles, resulting in higher temperatures at greater heights, independent of the precise form of $\gamma(T)$. This increase of temperature with height is shown qualitatively with an increase of the VDF broadness with height. Moreover, the temperature is the variance of the VDF normalized by the local density $n(z)$, then Fig. \ref{fig:VDFs} shows quantitatively this increase of temperature with height. We note that the figure is designed to show well the core distribution and since $\gamma_3(T)$ contained mostly coronal temperatures, the broad extension of its VDF (large $v$ values) is outside the plots.  This is why the VDF$/n(z)$ of $\gamma_3(T)$ is below the two others VDF$/n(z)$ at larger height while its $T(z)$ is larger. 

Finally, at coronal heights, the temperature continues to rise due to the persistence of suprathermal and leptokurtic velocity distribution functions, which maintain effective velocity filtration.

\section{Summary, discussion and perspectives}\label{sec4}

In this work, we have presented a kinetic model of the solar atmosphere in which the collisionless coronal plasma is in steady contact with a dynamically fluctuating chromosphere, modelled as a thermal boundary. Motivated by the routine observation of small-scale, transient brightenings on the Sun discussed in the introduction, the chromosphere is represented as a two-dimensional surface, with localized heating events randomly distributed across it. Given that the spatial extent of these events is much smaller than the solar surface, we have developed a surface coarse-graining procedure to describe the corona locally, by averaging over multiple events.

By performing this averaging over an intermediate surface $S$, sufficiently large to encompass many heating events but still small compared to the full solar surface, we have shown that the dynamics of a two-species (electrons and protons), collisionless plasma can be effectively described through coarse-grained distribution functions $\tilde{f}_{\alpha}$. These obey a set of Vlasov equations supplemented by non-thermal boundary conditions arising from the superposition of maxwellian at different temperatures.

We have derived analytical expressions for the stationary state distribution functions of both species, which depend solely on the single-particle energy $H_{\alpha}$. Within this framework, the observed anti-correlation between density and temperature naturally arises via the velocity filtration mechanism, in analogy to the original scenario proposed by \citet{Scudder1992a}. However, in contrast to that approach, suprathermal tails are not imposed a priori but are the consequence of 
spatial fluctuations in the chromospheric temperature (heating events). These fluctuations create a multi-temperature boundary condition, leading to non-thermal features in the stationary-state distribution functions.

Compared to previous work \citep{barbieri2023temperature,Barbieri2024b}, where temperature inversion was shown to result from temporal intermittency of heating at a fixed spatial location, the present model demonstrates that spatial intermittency, i.e., the inhomogeneous distribution of heating events across the base of the corona, is sufficient to produce similar non-Maxwellian stationary states. While both mechanisms lead to analogous inverted profiles and non-thermal distributions upon coarse-graining, the physical origin of variability (temporal vs spatial) is different, offering complementary insights.

In our model, the key parameter controlling the extent of heating, $A$, denotes the fraction of the surface heated to coronal temperatures ($\sim$1 MK). This parameter is calibrated to ensure that the average base temperature remains close to the chromospheric value, leading to a small $A$ consistent with an observed sparse distribution of heating events. While this qualitative picture aligns with current understanding, quantitative validation requires high-resolution solar observations in EUV, a direction we intend to pursue in future work using EUI and instruments on board of Solar Orbiter and AIA instrument on board of Solar Dynamic observatory. 

An important extension of this framework involves introducing a spatially varying magnetic field. In solar coronal structures such as loops, the magnetic field strength decreases with height due to flux-tube expansion \citep{Mandrini_2000}. Such variations, impose conservation of the magnetic moment, leading to anisotropic temperature profiles: the parallel temperature increases, while the perpendicular temperature decreases. This mechanism would enhance gravitational filtering along field lines. 

Although we model the coronal plasma as collisionless, real coronal conditions are not entirely free of collisions. However, because the collisional mean free path increases strongly with velocity ($\propto v^4$), low-energy particles are more affected, while suprathermal particles, responsible for reaching coronal heights, remain largely unaffected \citep{1983ApJ...266..339S,Landi-Pantellini2001}. Collisions would thus thermalize the low-energy part of the distribution while preserving the suprathermal tails, potentially making the velocity filtration mechanism even more efficient \citep{1983ApJ...266..339S,Landi-Pantellini2001}. This effect could lead to a sharper transition region. Studying the interplay between collisions and velocity filtration is a direction for future work.

Finally, we note that the temperature and density profiles predicted by the present model have the same analytical form as those derived in \citet{Barbieri2024b}, except for the difference in geometry: Cartesian here versus curvilinear there.
Since the model introduced in \citet{Barbieri2024b} was subsequently applied to low-mass main sequence stars \citep{barbieri2024temperaturedensityprofilescorona}, successfully predicting the observed temperature inversion in those systems, we expect that the present model, by construction, will yield the same conclusion.

\appendix

\begin{acknowledgements}
L.B. thanks Etienne Berriot and David Paipa-Leon for useful discussions. L.B. wants to thank
the Sorbonne Université in the framework of the Initiative Physique des Infinis for financial support.
\end{acknowledgements}

   \bibliographystyle{aa} 
   \bibliography{manuscript} 
\end{document}